\documentclass[11pt,twoside]{article}
\usepackage{asp2004}
\usepackage{psfig}
\usepackage{epsf}
\usepackage{graphics}
\usepackage{graphicx}
\usepackage{lscape}
\begin{document}
\title{How is the blazar GeV emission really produced?}
 \author{M. Georganopoulos$^{1,2}$, E. S.  Perlman$^{1}$, 
D. Kazanas$^{2}$, \& B. Wingert$^{1}$}
\affil{$^{1}$Joint Center for Astrophysics, University of Maryland, Baltimore County, 1000 Hilltop Circle, Baltimore, MD 21250}
\affil{$^{2}$NASA Goddard Space Flight Center, Code 661, MD 20771}

\begin{abstract}
We show that the  external Compton (EC)  model for the production
of the GeV emission in blazars makes specific predictions for the spectrum
and variability  of those blazars characterized by a high Compton 
dominance (Compton to synchrotron luminosity ratio). 
These unavoidable features
have not been observed, casting doubt on the validity of this 
popular model. We argue that synchrotron-self Compton (SSC) models
including the higher orders of Compton scattering 
are more promising, and we briefly discuss the implications of our findings 
for the geometry of the broad line region (BLR).
\end{abstract}

{\bf Introduction.}
The GeV emission from the relativistic jets of blazars is believed to be 
due to energetic electrons that inverse-Compton (IC) scatter to $\gamma$-ray 
energies  lower energy photons. 
These photons can be   synchrotron photons produced in the source 
(synchrotron-self Compton, SSC; e.g. Maraschi et al. 1993) and/or 
external photons such as
the BLR UV photons (Sikora et al. 1994) which, as reverberation mapping 
shows, are  produced at distances  of  $\sim 10^{17-18}$ cm from the central 
engine (Kaspi et al. 2000). 
The second case is believed to be favored,  because  variability arguments 
show that the site 
of the blazar emission is at a comparable distance, and, therefore, it is 
exposed to the BLR photon field, which has a  photon energy density in the jet  frame boosted by $\Gamma^2$, where $\Gamma\sim  10$ is the Lorentz factor of the 
jet bulk motion. 

\vspace{0.2cm}

{\bf Cooling in the Klein-Nishina (KN) regime.}
In several cases, the  Compton 
dominance can reach values as high as  a few hundred (e.g. PKS 4C 38.31; Kubo et al. 2000). 
In such cases IC is the dominant energy loss mechanism, and to produce the 
few  GeV photons with  energy  $\epsilon\sim 10^4$   
(in units of $m_ec^2$), electrons of at least the same Lorentz factor
are required, $\gamma \sim 10^4$. Given  that the typical BLR photon energy 
is $\epsilon_0\sim 10^{-4}$, the GeV emission comes from scatterings in the 
area between the Thomson and Klein-Nishina regimes, since  
$\epsilon_0 \gamma \sim 1$.

Consider a case where the only loss mechanism we have is IC.
In the Thomson regime ($\epsilon_0 \gamma \ll 1$), the electron energy loss 
rate $\dot\gamma\propto \gamma^2$, and the electron 
cooling time $\tau_c=\gamma/\dot\gamma\propto 1/\gamma$. 
In the KN regime $\dot\gamma\propto \gamma^0$ (there is a slow logarithmic increase of $\dot \gamma$ with $\gamma$ which we do not consider here), and  
$\tau_c=\gamma/\dot\gamma\propto \gamma$. So, 
while the cooling time decreases linearly with $\gamma$ in the Thomson regime,
it {\sl increases} linearly in the KN regime. 
The behavior around $\epsilon_0 \gamma \sim 1$ is flat, with a practically
energy independent cooling time, as a   numerical 
calculation (Fig. 1) shows. Assuming that a 
power law electron distribution $\propto \gamma^{-p}$ is injected
  in  the emission 
zone and that the electrons escape  after time $t_{esc}$, the steady-state
electron distribution $n(\gamma)$ will retain the same slope for $\gamma<\gamma_{b1}$, $\gamma>\gamma_{b2}$, where  $\gamma_{b1}$ and  $\gamma_{b2}$ are the
electron energies for which $\tau_c=t_{esc}$. To obtain $n(\gamma)$ for 
$\gamma_{b1} < \gamma < \gamma_{b2}$, we solve the steady-state
kinetic equation $\partial(\dot\gamma n)/\partial\gamma\propto \gamma^{-p}$
to obtain $n \propto \gamma^{1-p}/\dot\gamma$. In the Thomson regime
$\dot\gamma\propto \gamma^2$ and $n(\gamma)\propto\gamma^{-(p+1)}$, resulting
in  a distribution steeper than the injected one. 
Contrary to this well known result, in the KN regime
$\dot\gamma\propto \gamma^0$, and $n(\gamma)\propto\gamma^{-(p-1)}$, resulting
in an electron distribution {\sl flatter} than the injected one (e.g. Moderski et al. 2005). The general 
case is shown in the top panel of Fig. 1.
 Note that around  $\epsilon_0\gamma =1$, $n(\gamma)$ retains its initial
 slope as in the non-cooling parts  $\gamma<\gamma_{b1}$, $\gamma>\gamma_{b2}$, although these electrons are the fastest cooling.

\begin{figure}[t]
\begin{center}
\includegraphics[scale = 0.37, angle = 00]{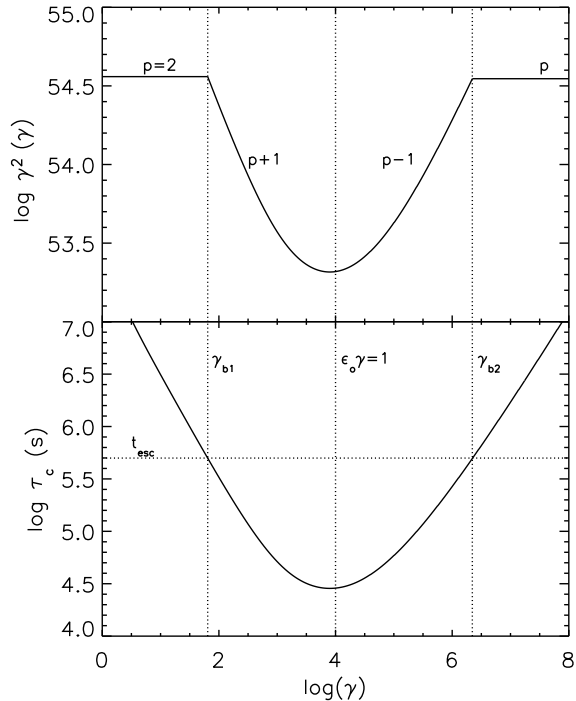}
\end{center}
\caption{\sl A homogeneous source. Bottom panel: the cooling time as a function of electron $\gamma$ for 
 electrons cooling due to IC scattering off a monoenergetic
photon field with $\epsilon_0 =10^{-4}$.  Only electrons
with $\gamma_{b1} < \gamma < \gamma_{b2}$ have time to cool. Top panel: the 
steady-state electron energy distribution $n(\gamma)$ 
(multiplied by $\gamma^2$ for visualization reasons) for an injected 
distribution $\propto \gamma^{-2}$.}
\label{fig1}
\end{figure}

\vspace{0.2cm}

{\bf Strongly Compton dominated blazars.}
In this case, a proper calculation
 requires the inclusion of the synchrotron and SSC losses, as well as the
 emission due to these processes and EC scattering. We present results 
 of such a  numerical calculation (Georganopoulos et al., in prep.) in Fig. 2,
 for a source in which the ratio of the
 energy density 
$U_{ext}$ in the jet comoving frame of the external (BLR) photon field  is 
100 times larger than the magnetic field energy density $U_B$.
This guarantees that the EC emission will be much more  powerful than 
the synchrotron component, while an adequately large source size 
($R=5 \times 10^{16}$ cm) guarantees that the SSC component 
is  much weaker than the EC one.

\begin{figure}[t]
\begin{center}
\includegraphics[scale = 0.44]{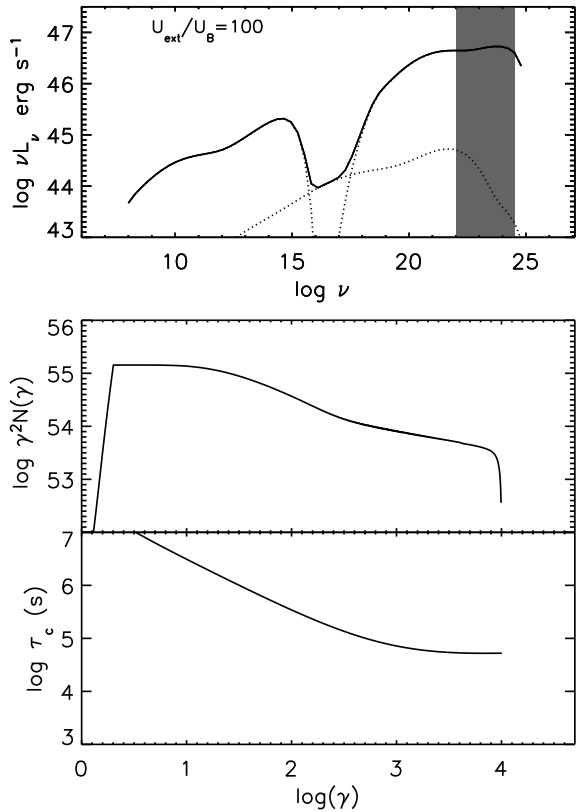}
\end{center}
\caption{\sl Model of an EC dominated blazar. Bottom panel: the electron cooling time as a function of $\gamma$. Middle panel: the electron distribution.Top panel:  the emitted power. Solid line for the total power, and dotted lines for the synchrotron (leftmost), SSC (central), and EC (rightmost and most powerful) components. The gray band is roughly the {\sl EGRET - GLAST} regime.}\label{fig2}
\end{figure}  

Note that the cooling time (bottom panel of Fig. 2) for electrons
with $\gamma$ greater than  $\sim 10^3$ is practically constant. 
{\sl This implies that the IR to UV  variability, produced by the 
synchrotron emission of these  electrons, should be achromatic,
contrary to observations (e.g. Ulrich et al. 1997, see however Perlman et al. 2003.} It also implies achromatic variations in the one to 
several GeV range, something that will be
 tested by {\sl GLAST}. Also, as can be seen in the middle panel of Fig. 2,
the electron distribution, after it softens due to cooling in the Thomson 
regime, it becomes harder due to the onset of the reduced efficiency 
KN cooling. This is reflected in the emitted spectrum with a  
hump in the synchrotron component, something that is also not observed 
in blazars (e.g. Kubo et al. 1998). 
Note finally the flat/rising   GeV component,  something rarely
 seen by {\sl EGRET} (the typical  GeV spectrum is steep; e.g. Kubo et al. 1998). 

{\sl The spectral and temporal characteristics we presented here are 
unavoidable for blazars with high Compton dominance, and the fact that 
these characteristics are not seen is a strong argument against the EC model. }

\vspace{0.2 cm}

\begin{figure}[t]
\begin{center}
\includegraphics[scale = 0.5, angle = 00]{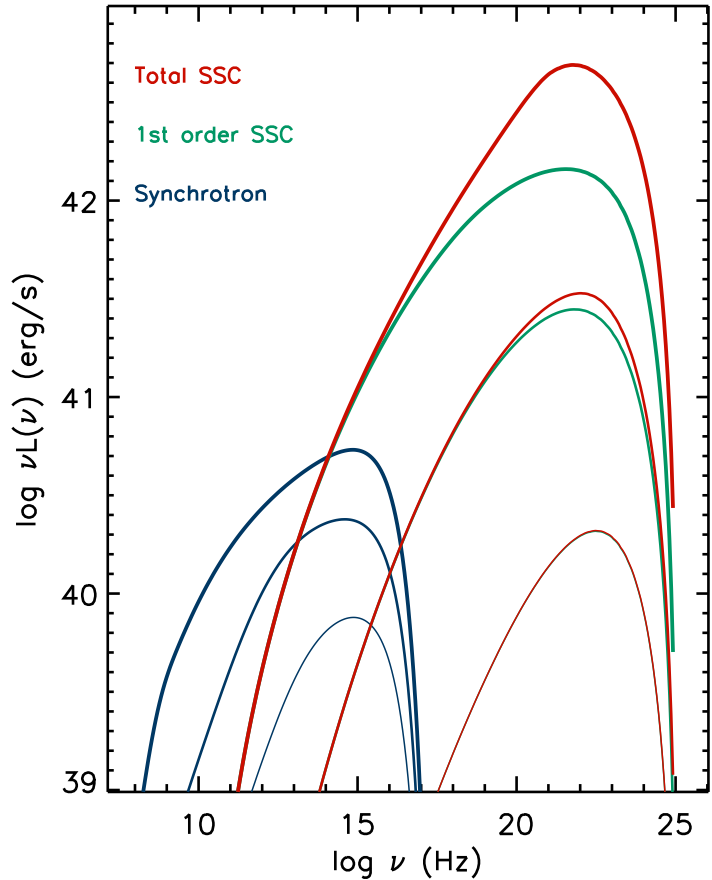}
\end{center}
\caption{\sl In this source, a variation of the injected electron power by
10 and 100, resulted to synchrotron variations by 3.2 and 7.1, and to
SSC variations by 16 and 234. The superquadratic behavior of the source is partially due to the onset of the second order (SSC2) scattering.} \label{fig3}
\end{figure}

{\bf Revisiting SSC.} Given the problems of the EC interpretation of the
 GeV emission, we turn to the SSC interpretation. One of the main 
 reasons that SSC 
models for the blazar GeV emission were disfavored, was
an analytical  argument that SSC variability is quadratic as compared with 
that seen in the synchrotron component,  contrary to the 
  superquadratic variations  seen in 3C 279 (Wehrle et al. 1998). 
Interestingly, the currently favored EC mechanism 
can only produce linear  variations in a simple fashion, 
and modeling superquadratic variations requires
a carefully selected  change  of more than one  model parameter.

The argument for the solely quadratic SSC variations was based on  the 
 assumption that the  SSC power is much smaller than the synchrotron 
one. In the more general and observationally relevant case of an SSC power
comparable or higher  than the synchrotron one,  superquadratic variations are 
the norm, particularly so  when second order (SSC2) scattering is relevant 
(see Fig. 3).

\vspace{0.2 cm}

{\bf The BLR geometry.} Given that the blazar emission site cannot be placed
much further out than $10^{18}$ cm from the central engine, it is interesting
to ask under what conditions the photon field of the BLR is not an important
seed photon mechanism for IC emission. Reverberation mapping provides
us with a typical size of the BLR region (for a powerful blazar like 3C 279, this is  $\sim 1.5 \times 10^{17}$ cm, following Kaspi et al. 2000), 
but {\sl not} with its geometry.
If the BLR has a flattened geometry  (see Fig. 4) with radius $\sim  10^{17}$ cm,  then at a distance of $\sim 10^{18} $ cm from the central engine, 
where the blazar emission site lies, the BLR photons illuminate the blazar from behind, and their comoving energy density scales as $1/\Gamma^2$, instead of 
$\Gamma^2$ for the case where the BLR enclosed the blazar site. This results to
a strong decrease of the EC emission.
We note that several independent arguments (e.g. Wills \& Browne 1986, Maiolino et al. 2001, Rokaki et al. 2003) for a flattened 
BLR geometry  have been advanced in the literature, making this scenario
very plausible. 

\vspace{0.2cm}

{\sl GLAST observations of high Compton dominance blazars 
will be critical in establishing this new picture that not only address the 
mechanism of the GeV emission, but also provide constraints
 on the spatial distribution of
the gas clouds close to the central engine. }

\begin{figure}[t]
\begin{center}
\includegraphics[scale = 0.4, angle = 00]{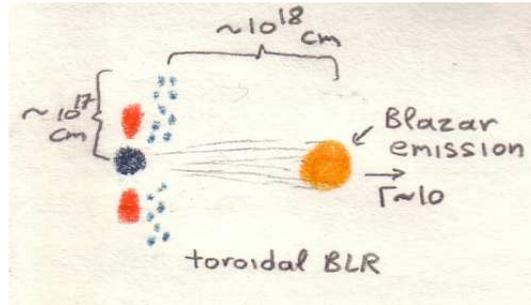}
\end{center}
\caption{\sl A pita - like BLR on top of the accretion disk with a size
$\sim 10^{17}$ cm and a blazar emission site at $\sim 10^{18}$ cm. 
In this geometry, the BLR comoving energy density drops by up to 
 $\; \Gamma^4$ relative to the case of a spherical or shell-like BLR 
that encloses the blazar emission site.}\label{fig4}
\end{figure}

\end{document}